\documentclass[12pt]{iopart}
\usepackage{bm,amsfonts,amssymb, graphicx}
\usepackage{epstopdf}

\begin{document}

\title{IR-Laser Welding and Ablation of Biotissue Stained with Metal
Nanoparticles}
\author{A. A. Lalayan, S. S. Israelyan}

\address{Centre of Strong Fields Physics, Yerevan State University, 1 A. Manukian,
Yerevan 0025, Armenia} 
\ead{alalayan@ysu.am}

\begin{abstract}
In the present work we have studied the possibility of laser welding and
ablation of biological tissue by the using of spherical metal nanoparticles
(NPs) and infrared laser irradiation which spectrally located far from
plasmon resonances. YAG:Nd laser with 1064 nm wavelength, 8ns pulse duration,
and operating in transverse electromagnetic modes TEM$_{00}$ was used for
the synthesis of metal NPs. The Au,Ti Ni and Cu as well as Au-Ag and Au-Cu
hybrid metal NPs were formed in the liquid medium. Effectiveness of laser
ablation in the case of the biotissue sample that stained with the metal NPs
was approximately on 4-5 times larger than for the native sample. Also the
scheme of a laser point welding for the deep-located biotissue layer
selectively stained by the metal NPs has been demonstrated.
\end{abstract}



\vspace{2pc} 

\maketitle
\section{Introduction}

Nanoparticles (NPs) are homogeneous and composite materials with size in a
range of 1-100 nm which exhibit some unexpected optical, electromagnetic,
chemical and mechanical properties not inherent to bulk materials and
atomic-molecular structures due to quantum confined nature of their energy
levels and surface area to volume ratio. The unique properties of NPs depend
on their size, morphology, composition, uniformity, and agglomeration. These
extraordinary qualities make superior the application of NPs in
nanophotonics, biomedicine, environment science, engineering etc \cite{1}-\cite{5}.
Especially, in medical applications the NPs are able to pass through
biological filters inside of organism by that deliver drugs. Also,
nanomaterials are used for protein detection, fluorescent biological labels,
tumor destruction, tissue engineering etc. Various physical and chemical
processes are currently widely used for the production of NPs, nevertheless
we focus on the laser synthesis of colloidal spherical NPs by the laser
ablation in liquid media \cite{6}-\cite{9}. This method favorably differs from
other methods in its simplicity, efficiency, and it does not harm the
environment. Also, by this method we can obtain nanoparticles of different
sizes varying in the wide range and strongly depending on the details of
experimental setup. In work the \cite{6} we have reported on laser synthesis
of GaAs and CdSe semiconductor NPs with picosecond Nd:YAG laser. The
considerable blue shift of the photoluminescence was observed when
picosecond laser beam with transverse electromagnetic modes TEM$_{00}$
structure has been used. Particle sizes was estimated (2-3 nm) from the
spectral data and the bandwidth of the luminescence spectra of about 40 nm
reveals a narrow size distribution of the produced quantum dots which was
achieved without any size selection methods. Such ultra fine sizes of NPs
are excellently meet with the requirements for medical applications. In
recent years the use of nanoparticles in biomedicine for the early
detection, accurate diagnosis and treatment of cancer is extensively
studied. Particularly, the possibilities for cancer photothermal therapy due
to tissue heating by using metal NPs have been successfully demonstrated 
\cite{10,11,12}. This method is based on localized surface plasmon resonance
(LSPR), when nanoscale localization and amplification of electromagnetic
fields occur in the vicinity of metal nanoparticles \cite{13}. The locally
amplified optical fields generate localized thermal energy due to the rapid
conversion of photon energy into heat via electron--electron scattering and
electron--phonon coupling \cite{14} in the wavelengths of the plasmon
resonance \cite{15}-\cite{18}. The plasmon resonances for gold and silver
nanospheres are in the green-blue range of the visible spectrum, which can
be red-shifted into the near infrared if nanoparticles shape is modified to
nanorods or nanoshels. This is more practical as biological tissues are
relatively transparent to near infrared light. However the strong tuning to
plasmon resonance wavelength is required in medical applications of
photoheating that in most cases is not provided by commercial medical laser
devises.

\section{Experimental results}

For syntheses of metal NPs in the considered experiment, the YAG:Nd laser
with 1064nm wavelength, 8 ps pulse duration, repetition rate of 10Hz and
operating in transverse electromagnetic modes TEM$_{00}$ was used. The same
scheme of laser synthesis described in the work \cite{6} has been used. The
laser beam was focused on the surface of a bulk metal target allocated in
the glass cuvette with distilled water. Exposition of the laser irradiation
was two hours. The Au, Ti, Ni and Cu, as well as Au-Ag and Au-Cu hybrid
spherical metal NPs were formed in the liquid medium. Laser welding and
ablation of biotissue stained with the metal nanoparticles has been studied
on samples of chicken skin. Figure (1) shows typical photographical
picture (taken by an optical microscope) of tissue stained with the metal
NPs. 
\begin{figure}[tbp]
\begin{center}
\includegraphics[width=.5\textwidth]{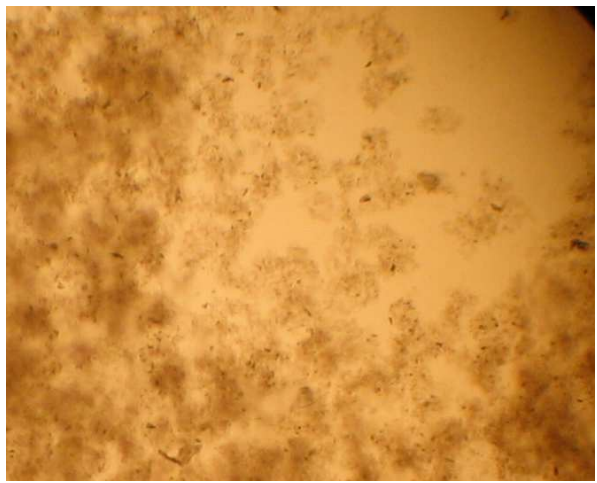}\label{fig1}
\end{center}
\caption{Picture of chicken skin tissue sample stained with the metal NPs.}
\end{figure}
In the present work we have studied the possibility of tissue heating by
using infrared laser irradiation at 1064 nm wavelengths. This wavelength is
located far from plasmon resonances of known spherical metal NPs.
Particularly, silver and gold nanoparticles exhibit strong plasmon resonance
at wavelengths 405 nm and 520 nm correspondingly. The absorption properties
of colloidal solutions of Ag, Au and Cu nanoparticles that were synthesized
by method of laser ablation in a liquid has been measured (see Figure (2)). 
Obtained data are consistent with the previous publications on
plasmon resonances of widely used Ag, Au spherical NPs with the sizes of
about 50 nm. 
\begin{figure}[tbp]
\begin{center}
\includegraphics[width=.6\textwidth]{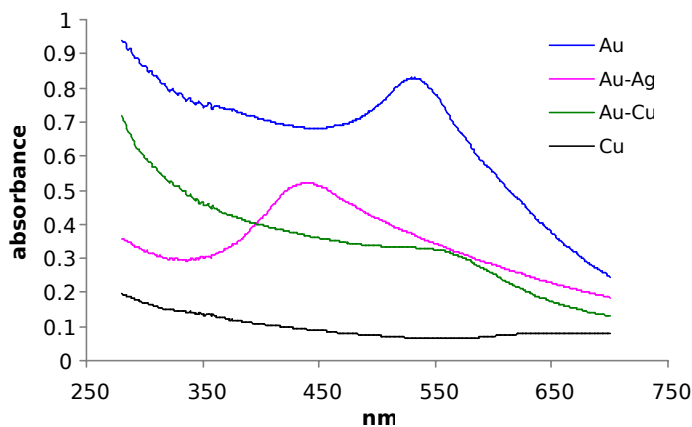} \label{fig2}
\end{center}
\caption{Absorbance spectra of Au, Cu, Au-Ag and Au-Cu nanoparticles.}
\end{figure}
As we can see in Figure (2), the plasmon resonance of Cu
nanoparticles in comparison with other considered metal NPs is located in
the red range of the optical spectrum, however, its value is rather smaller.
The continuous wave YAG:Nd laser with an output beam of 3mm diameter and
power up to 4W was used for the biotissue welding and ablation. Two areas of
sample's surface - unstained and stained with metal nanoparticles were
ablated at the same dose of the laser irradiation. Figure (3) shows
the picture of skin surface after ablation procedure. 
\begin{figure}[tbp]
\begin{center}
\includegraphics[width=.8\textwidth]{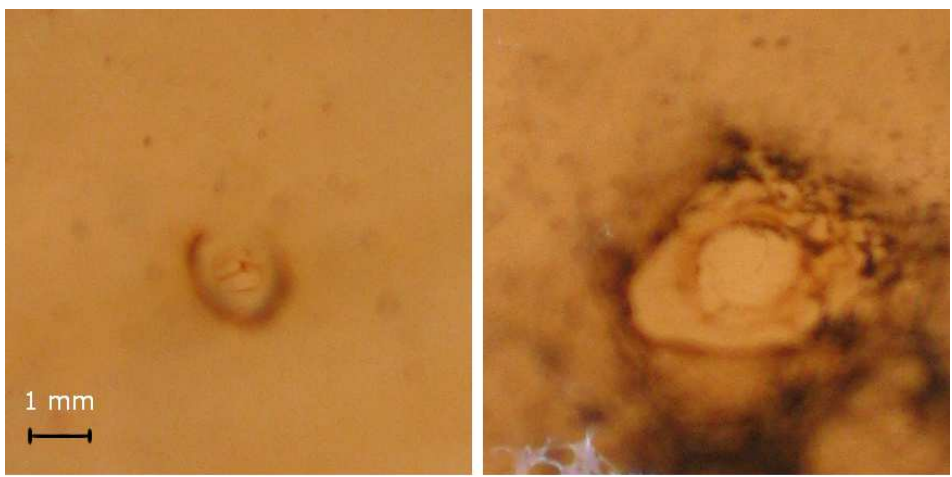} \label{fig3}
\end{center}
\caption{Images of sample's surface after laser ablation procedure. The left
side of the picture represents non-colored area of the sample and the right
side -- the sample colored with silver (Ag) nanoparticles.}
\end{figure}
The surface of the area which has been exposed to laser ablation in the case
of the tissue sample that colored with silver NPs is approximately on 4-5
times larger than that for the sample with non-colored area. Furthermore,
the same result was obtained in case of several other metal nanoparticles
with different wavelength of plasmon resonances, as well as with Ni and Ti
nanoparticles which does not exhibit any plasmon resonance properties. Thus,
we observe significant difference in photodamage for biotissue samples which
are unstained or stained with the metal nanoparticles. Such difference in
absorption of infrared radiation may be explained by the presence of the
metal nanoparticles clusters in the tissue samples. Indeed, as we can see in
the Figure (1), the coloration of biotissue is highly
nonhomogeneous and one can observe the number of cluster centers colored in
intense black. In the book \cite{19} there are given experimental and
numerical analysis of optical characteristics of fractal aggregates of
nanoparticles. As it is shown, the absorption spectra for clusters with
large amount of nanoparticles (500-10000) are shifted toward infrared range.
This result demonstrates that the application of metal NPs in laser surgical
procedures will lead to significant reduction of an irradiation dose even in
cases when plasmonic resonance is absent or the wavelength of laser
radiation is far from it. Furthermore the application of nanoparticles in
laser surgery allows effectively use IR laser light that penetrates deeply
in the tissue and absorbed poorly by the native tissue itself. However, this
radiation may be absorbed strongly by the nanoparticles aggregates. This
advantage can be used for development of the schemes of laser welding for
deep-located biotissue areas selectively colored by the metal NPs. For
example, we performed the point welding of two chicken skin samples with
thickness of 2 mm each (see Figure (4)). 
\begin{figure}[tbp]
\begin{center}
\includegraphics[width=.4\textwidth]{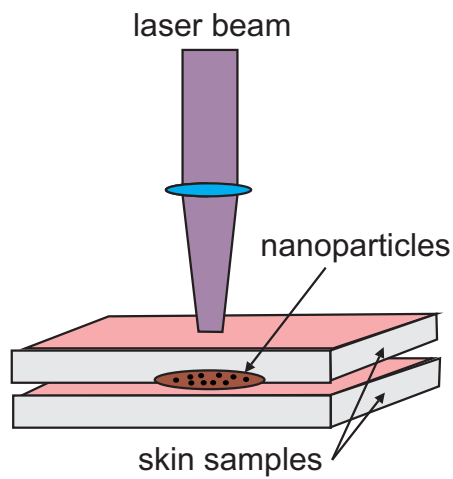} \label{fig4}
\end{center}
\caption{Scheme of laser point welding of two skin samples colored with the
silver nanoparticles.}
\end{figure}
The surfaces of the samples were colored with the silver nanoparticles. The
laser radiation penetrates through the first sample from the noncolored side
without production of the visible damages of tissue and reaches the
contacting area of samples that contain nanoparticles. In this case we
observe the strong absorption of light in the area of biotissue colored with
the nanoparticles and, subsequently, the local heating of the tissue. In the
result, the local spot welding of two layers of the skin tissue was
realized. Note that in case of the using of radiations of 405 nm or 520 nm
wavelengths which are in the range of Ag and Au plasmon resonances, we would
have strong absorbance by the tissue of 2 mm thickness, which will lead to
the photodamage of the superficial layers. Furthermore, the application of
Ag nanoparticles that exhibit strong microbicidal effect may lead to
significant improvement of the postsurgical healing of the patients. It is
also important as tissue photoheating could cause inflammatory process due
to local antimicrobial immune resistance reduction, and this negative effect
could be also minimized.

\section{Acknowledgments}

We would like to thank Prof. H. K. Avetissian for valuable discussions. This
work was supported by State Committee of Science of RA.

\end{document}